\documentstyle[twocolumn,aps]{revtex}
\begin{document} 
\draft
\title{Quantum gates by coupled asymmetric quantum dots  
and controlled-NOT-gate operation}
\author{Tetsufumi Tanamoto}
\address{Corporate Research and Development Center, 
Toshiba Corporation, 
Saiwai-ku, Kawasaki 210-8582, Japan}
\date{\today}
\maketitle

\begin{abstract}
A quantum computer based on an asymmetric 
coupled dot system 
has been proposed and shown to operate as 
the controlled-NOT-gate.  
The basic idea is (1) the electron is localized 
in one of the asymmetric coupled dots. 
(2)The electron transfer takes place from one 
dot to the other when the energy-levels of the 
coupled dots are set close. (3)The Coulomb interaction 
between the coupled dots mutually affects the energy 
levels of the other coupled dots. 
The decoherence time of the quantum computation 
and the measurement time are estimated.
The proposed system can be realized 
by developing the technology 
of the single-electron memory using Si nanocrystals 
and the direct combination of the quantum circuit and 
the conventional circuit is possible. 
\end{abstract}

\bigskip
\narrowtext

\section{Introduction}
Since Shor's factorization program was proposed, 
many studies have been carried out with a view to 
realizing the quantum computer
\cite{Ekert,Gershenfeld,Barenco95,DiVincenzo98,Kane98,Schon,Nakamura}. 
Although coherence is necessary for a quantum calculation, 
it is considered to be difficult to maintain coherence 
in the entire calculation process 
throughout the entire circuit. 
Thus, it will be more efficient and more realistic to combine 
the quantum computational circuit and the conventional 
LSI circuit in the same chip. 
Some proposals regarding the quantum computer based 
on semiconductor physics have been made from this 
viewpoint\cite{Barenco95,DiVincenzo98,Kane98}. 

Kane\cite{Kane98} proposed the Si-based quantum 
dot computer using NMR of dopants (phosphorus). 
This idea is very promising because the qubits 
are isolated from the external environment which causes 
decoherence. However, controlling  the implantation of 
phosphorus exactly into the definite positions of the Si substrate 
will depend on future technology and the usage 
of the magnetic field seems to be undesirable in the 
 conventional Si LSI circuit. 
Here, we propose a coupled quantum dot system 
of a quantum computer which can be operated 
only by electrical effects and show that it 
can operate as a controlled-NOT-gate. 
It is shown to be realized by developing the 
technology of a single-electron memory of 
Si nanocrystals\cite{Tiwari,Guo}. 

The controlled-NOT operation is given by\cite{Barenco95}:
$|\epsilon_1 \rangle |\epsilon_2 \rangle 
\rightarrow 
|\epsilon_1 \rangle |\epsilon_1 \oplus \epsilon_2 \rangle  
({\rm modulo 2}) $
where $\epsilon_1$ shows a {\it control qubit } and 
$\epsilon_2$ shows a {\it target qubit}. 
The value of $\epsilon_1$ remains unchanged whereas 
that of $\epsilon_2$ is changed only if $\epsilon_1=1$. 
This operation is important  
because it acts as a measurement gate and  produces the 
entanglement\cite{Barenco95} 
which plays an important role in 
quantum cryptography gates\cite{Bennett}. 
In this paper we show the quantum gates of 
the semiconductor coupled quantum dots, 
emphasizing their controlled-NOT operation.

The structure of the paper is as follows. 
In Sec.\ref{sec:c_not} 
the basic idea of this paper is presented 
and the static and dynamic properties 
of the coupled quantum dots operating as a quantum gate 
are discussed. 
In Sec.\ref{sec:discussion} we 
estimate the decoherence time  
in the quantum operation and the measurement time 
in the detection process 
of the proposed coupled quantum dot system. 
We also discuss the fabrication process 
of the coupled quantum dot system. 
Conclusions are presented in Sec. \ref{sec:conclusion}. 

\section{Controlled-NOT-gate by the two coupled dots} 
\label{sec:c_not}
Coupled quantum dot systems with a few electrons 
have been  extensively investigated 
in regard to many-body effects such as Coulomb blockade
\cite{Stafford94,Weiss,Vaart,Pfannkuche,Crouch,Waugh}. 
From the experiments by van der Vaart\cite{Vaart},
it can be seen that the electron transfer between dots 
occurs when the discrete energy-level of one of the dots  
matches that of the other dot of the coupled dots.
Pfannkuche {\it et al.}\cite{Pfannkuche} showed 
theoretically that, as a result of the correlations 
between a few electrons in quantum dots, 
the electrons behave as if they were noninteracting electrons. 
Crouch {\it et al.}\cite{Crouch,Waugh}
showed that, if the tunneling barrier is low and 
the coupling of the two dots is strong, 
the coupled dots behave as a large single dot 
in a Coulomb blockade phenomenon. 
This means that, if the tunneling barrier between the dots 
is sufficiently small, it is possible that 
only one electron exists in the coupled dots. 

Thus, we can consider the electronic state of the 
two coupled dots in the range of the free 
electron approximation\cite{Yariv,Tsukada} at the first 
step of investigation.  
When two dots of different size are coupled
and one excess electron is inserted, 
the system can be treated as a two-state system 
where the energy-levels of the total coupled dot 
system show the localized state of 
wave function reflecting the different energy-levels 
of the independent isolated dots\cite{Yariv}.  
When gate bias voltage is applied and the potential 
slope is changed, there appears a gate bias voltage, $V_{\rm res}$, 
at which the two energy-levels of the original single 
dots coincide and 
the electron transfers to another dot (resonant tunneling). 
Coupling removes the degeneracy of energy-levels in 
the single dot quantum states, and produces new states 
of delocalization such that the even-parity and odd-parity wave 
functions spread over the two coupled dots. 
Thus, if we regard the perfect localization of the 
charge in one of the coupled dots as '$|1 \rangle$' state 
and that in the other dot as '$|0 \rangle$' state, 
we can constitute {\it qubit } by the coupled quantum dots
 (Fig. \ref{fig1}). %

The point is, by adjusting the gate bias, 
we can control the electronic state from the 
localized regions ( $ |1 \rangle $ and $ |0 \rangle $ ) 
to the intermediate delocalized region only where 
the electron transfers from one dot to the other 
in the coupled dots in a short time 
($\sim$ picosecond)\cite{Yariv,Tsukada}. 
As the tunneling barrier structure is asymmetric, 
the leak current through the coupled dot system 
is extremely small\cite{Ricco} and neglected actually.

When the above coupled dots (qubits) are arrayed side by side, 
the charge distribution of the electron in a qubit 
changes the potential profiles 
and the energy-levels of the neighboring 
qubits by its  electric field. 
We assume that the electron transfer between different 
qubits can be neglected. 
Then the electronic state in a qubit is affected 
by whether the electrons in other qubits 
stay in  the '$|1 \rangle$' state, 
 '$|0 \rangle$' state or  arbitrary superposition state 
of '$|1 \rangle$' and '$|0 \rangle$'. 
By changing the charge distribution of the 
array of qubits, we can operate the 
total charge distribution of the electrons 
and the quantum circuit. 
Figure \ref{fig2} shows the case of the two-qubit controlled-NOT-gate 
where one set of the coupled dots operates 
as a {\it control qubit} and the other as a 
{\it target qubit}. 
Later, it is shown numerically that this array of the coupled 
quantum dot operates as the controlled-NOT-gate.  

One of the candidates for the coupled dots of the quantum 
computer is considered to be the Si nanocrystals 
embedded in the gate insulator (Fig.\ref{example}).
This is based on the Si LSI technology similar to that of 
Tiwari's single-electron memory\cite{Tiwari,Guo}
which is extensively investigated 
because it operates at room temperature. 
The excess charge is supplied  
from the inversion layer in the substrate. 
By setting larger dots near the channel, 
the structure shown in Fig. \ref{fig2} 
can  be realized. 
The arrangement of the gate electrodes 
which control the individual qubits 
depends on the individual algorithm. 
The simplest form is the case where there are  two 
gate electrodes and two sets of coupled dots (Fig. \ref{fig2}), 
which works as the controlled-NOT-gate explained in this paper.
The {\it measurements} process is operated by 
the upper gate electrode which controls the overall 
channel carrier density. The upper gate also protects 
the electronic states in the dots from 
disturbance by shielding the external electromagnetic field. 
The qubits interact mutually and the distribution 
of the charges affects the current flow (channel conductance)  
between the source and drain  
and the threshold gate voltage. 
A $|11 \rangle $ state shifts the threshold voltage most and 
a $|00 \rangle $ state shifts it least. 
Because of the sloping channel depth from the source to the drain, 
the  $|10 \rangle $ state and $|01 \rangle $ state 
can be distinguished. 
Thus the quantum mechanical calculation proceeds as follows:
(0)To {\it initialize } the charge distribution 
(initial quantum states), a large voltage 
is applied on the upper gate 
over the coupled dots, and unifies the charge 
distribution in the coupled dots,
(1)The input and output signals are added through the 
gates over each qubit,
(2)the final distribution of charges (final quantum states) 
is detected by the current between the source and drain 
and the threshold voltage shifts of the upper gates 
over the coupled dot system.

When there are many qubits, the controlled-NOT operation of pairs of 
qubits is affected by the quantum states of the surrounding qubits. 
That is, the applied gate voltage of operation changes depending on 
the quantum states of other qubits. 
This is the same situation as the qubits of Barenco {\it et al}.\cite{Barenco95}. 
Although the decoupling schemes used in NMR experiments can be 
applied to avoid the coupling between qubits, it is considered to be desirable that the general quantum 
calculations are designed by considering the arrangement of 
qubits\cite{Lloyd}.

This structure of the proposed system has the merit that 
the charge distribution in the coupled dot system, 
which is considered to be a very small signal, 
is expected to be detected by the channel conductance 
with high sensitivity like that of the single-electron 
memory\cite{Tiwari}. 

Similar to Tiwari's single-electron memory, 
the charging effect appears between the 
coupled dots and the channel region
and the probability that two electrons 
come into the qubits is very small as long as 
the capacitance of the junction is sufficiently small. 

In principle, the qubits by the semiconductor quantum dots with discrete 
energy-levels do not directly require that the quantum dots  be asymmetric. 
Two coupled quantum dots with discrete energy-levels can organize 
the two-state system. However, when the coupled dots are embedded 
in the FET-type insulating layer that we propose, the asymmetry of 
the coupled dots is required for the following two reasons. 
First, in order to prevent the electron in the coupled dots 
from returning to channel region in the substrate during the quantum 
operation, a finite voltage is required. The second reason is related to 
the measurement process. 
For the current to flow, a finite gate voltage that is larger than 
the threshold voltage is required. Because, as discussed below, a 
large gate voltage breaks the coherent state of the coupled dots, 
much voltage cannot be applied on the gate electrode. Thus, it is 
desirable that the two discrete energy-levels of quantum dots 
coincide under the applied gate voltage and the quantum calculation 
and the measurement be carried out near the threshold gate voltage. 
These are the reasons why the asymmetry of the coupled quantum dots 
is required. 

Below, we show the  static properties 
of the wave function of the localized electron 
by using the  S-matrix theory 
and the controlled-NOT operation of the coupled dots. 
The periodical motion of the localized electron 
is shown by solving a time-dependent Schr\"{o}dinger equation.  
The exact theoretical treatment of the coupled dot system would be 
to solve the exact three-dimensional Schr\"{o}dinger equation. However, 
since this direct method is difficult to apply in practice, we use the 
following approximations. The static behavior is studied by solving 
the one-dimensional Schr\"{o}dinger equation, and the dynamic behavior, 
which is more difficult to treat, is studied by regarding  the quantum 
dots as zero-dimensional objects. 

\subsection{Static properties of the qubit of the 
coupled quantum dots} 
The static properties of the wave function in 
the coupled dots can be shown by 
applying the S-matrix theory\cite{TsuEsaki} 
to the one-dimensional case.
The one-dimensional Schr\"{o}dinger equation is given:
\begin{eqnarray}
[-\frac{\hbar^2}{2m_i} \frac{\partial^2 }{\partial z^2}
+V_i (z) ]\psi_{i} (z) =E \psi_{i} (z)
\end{eqnarray}
where $i(=1,..,N_{\rm m})$ show the number of the mesh in the calculation. 
It is well known that a relatively small number of $N_{\rm m}$ is sufficient 
for the calculation (here $N_{\rm m}$ $\sim$ 1000). 
The electric fields by other coupled dots 
are considered to be included in the potential, 
$V_i (z)$. 
We use the plane wave approximation for the 
wave functions:
\begin{equation}
\psi_i (x)=A_i e^{ik_i x} +B_i e^{-ik_i x}
\end{equation}
The boundary conditions are given  
\begin{equation}
\psi_i (x) = \psi_{i+1} (x), \ \ \
\frac{1}{m_i}\frac{\partial \psi_i}{\partial x}
=\frac{1}{m_{i+1}}
\frac{\partial \psi_{i+1}}{\partial x}
\end{equation}
which determines the coefficient $A_i$, $B_i$ :
\begin{equation}
\left[
\begin{array}{c}
A_{i+1} \\  B_{i+1} 
\end{array}
\right]=
\left[
\begin{array}{cc}
(1+r_i)e^{ik_i x} &  (1-r_i)e^{-ik_i x} \\
(1-r_i)e^{ik_i x} &  (1+r_i)e^{-ik_i x} \\
\end{array}
\right]
\left[
\begin{array}{c}
A_{i} \\  B_{i} 
\end{array}
\right]
\end{equation}
where $r_i = (k_i/m_{i}^* )/(k_{i+1}/m_{i+1}^*)$. 
Here we assume that the electron is inserted from 
the channel layer ($i=0$ part) and neglect the reflection 
amplitude of the wave function of the gate electrode 
($B_{N_{\rm m}}=0$ ), similar to Ref.\cite{TsuEsaki}. 
Then the transmission coefficient is given:
\begin{equation}
T_{N_{\rm m}}(E)\equiv \frac{|k_{N_{\rm m}}|}{|k_0|}
|A_{N_{\rm m}} |^2. \label{eqn:tr}
\end{equation}
Discrete energy-levels of the coupled dots 
are those when 
this transmission coefficient has a maximum. 
We can estimate the effects of the Coulomb interaction 
of the control qubit on the target qubit shown in Fig. \ref{fig2}. 
The Coulomb interaction on the dot $a_1$ 
from the dot $a_2$,  and that from 
the dot $b_2$ are given by 
$U_{a_1a_2}=e^2/\epsilon r_{a_1a_2}\rho_{a_2}$,
$U_{a_1b_2}=e^2/\epsilon r_{a_1b_2}\rho_{b_2}$, respectively, 
where  $\rho_i$ is the density of the wave function 
of dot $i$ and $r_{ij}$ shows the distance 
between the center of the dot $i$ and $j$. 
We set  $\rho_i=0$ or 1 depending on 
the existence of the localized electron 
of the neighboring qubits. 
The Coulomb interaction on the dot $b_1$ is treated similarly. 
These Coulomb interactions are added to 
the potential bottom of the target qubit.
For simplicity, we neglect the 
self-consistent effects.  

The localized electron in one of the coupled dots 
moves into the other dot only if the 
two discrete energy-levels are set close (on resonance). 
The slight change of the relative energy-level 
by the electric field generated by 
the other set of coupled dots 
makes impossible the transfer of the localized electron 
from one dot to the other. 
The basic concept of this scheme is similar 
to that of Barenco {\it et al}.\cite{Barenco95}.
Whereas Barenco {\it et al}. use the ground state 
and excited state in a single dot with optical resonant effects, 
we are using only the ground state 
of the coupled dots only with electrical resonant effects
({\it  ground state operation }). 
The smaller the size of the dot, 
the more stable is the operation. 

Figures \ref{fig4} and \ref{fig5} show the calculated results of the 
controlled-NOT operation in Si/SiO${}_2$($\epsilon=4$) material. 
The barrier height of SiO${}_2$ is assumed to be 3.1 eV and 
the effective mass of Si and SiO${}_2$ are assumed to be
$0.2m_0$ where $m_0$ is a mass of free electron. 
As the tunneling barrier is sufficiently high, 
the coupled dot system can be made smaller in the 
Si/SiO${}_2$ system than in the GaAs/AlGaAs system\cite{Tsu3}. 
The diameter of the larger quantum dot is 6nm and 
that of the smaller is 4nm where the thickness of the 
tunneling barrier is 1.5nm. 
The distance between centers of the dots of the same size 
is assumed to be  20nm (4$\times$ 10${}^{12}$ dots/cm${}^{2}$). 
These values are 
taken from the experiments by Tiwari {\it et al}.\cite{Tiwari}.
The thinner tunneling barrier between the dots in a qubit 
becomes ($\leq$ 1nm), 
the weaker the rate of the localization effect is. 

Electron in a target qubit is localized in a larger dot 
at lower gate bias region (region $A^{(1)}$ in the case 
where control qubit is in $|1 \rangle$ state in Fig.\ref{fig4}). 
As the gate voltage is applied, the energy-level 
of the localization in the larger dot 
exceeds that of the smaller dot. 
At $V_{\rm G}=V_{\rm res}^{(1)}$ (center of the 
left hatched region in Fig.\ref{fig4}), 
the wave function of the lowest energy state (even-parity) 
and that of the excited state (odd-parity) spread over 
the two dots with equal weight, 
when the control qubit is in $|1\rangle $ state. 
The degenerate energy-levels of the single dots are 
split by the coupling of the dots 
and show the small energy difference, $\Delta E$ ($\sim$ 6.28 
$\times 10^{-5}$ eV: 
which is not distinguishable in the figure).
This resonant gate bias  shifts toward the 
higher bias region in the case where the control 
bit is in $|0 \rangle$ state. This is because the 
electron in the control qubit is localized in a 
smaller dot and the band bottom of the smaller dot 
in the target qubit is raised (Fig.\ref{fig5}(b)).  
Thus, when we apply the voltage, $V_{\rm res}^{(1)}$, 
in a half time of the oscillation 
with time period, $\tau_{\rm \delta}^C \sim \hbar/2\Delta E$ 
($\sim 5.2$ picosec),  
the electron moves between alternate dots only if the 
control qubit is in  $|1 \rangle$ state and 
we can show the controlled-NOT operation 
in the coupled quantum dot system. 

\subsection{Dynamic properties of the qubit of the 
coupled quantum dots} 
Note that the wave function shown in Fig.\ref{fig4} is 
the static one and dynamical properties can be easily 
discussed by solving the time-dependent 
Schr\"{o}dinger equation\cite{Tsukada} as follows. 
The localized wave functions in a quantum dot $a$ and $b$ and 
the eigenenergies are 
expressed as $\psi_a (x)$ and $\psi_b (x)$, and,  
$E_a$ and $E_b$, respectively,
 which are assumed to be far apart from each other. 
Then the coupled wave function is constituted by these 
wave functions as, 
\begin{equation}
\psi(x,t) = a(t) \psi_a (x) + b(t) \psi_b (x).
\end{equation}
The Hamiltonian of the Schr\"{o}dinger equation, 
$i\hbar \partial \psi/\partial t= H \psi$, is given by 
\begin{equation}
H=-\frac{\hbar^2}{2m} \frac{\partial^2 }{\partial x^2} 
+V_a (x) + V_b (x) -V_0, 
\end{equation}
where $V_0$(=3.1 eV) is a barrier height between the 
quantum dot $a$ and $b$. 
This equation is easily solved and 
the eigenenergies are given as 
\begin{equation}
\omega_{\pm} = ( \omega_a + \omega_b )/2 \pm \omega_0 
\end{equation}
where $\omega_a = E_a/ \hbar$, $\omega_b=E_b /\hbar$, 
$c_a= \langle 1| V_a-V_0 |2 \rangle /\hbar$,   
$c_b= \langle 2| V_b-V_0 |1 \rangle/\hbar$  and 
$\omega_0 =$  $\sqrt{[(\omega_a -\omega_b)/2]^2 + c_a c_b}$. 
When the applied bias puts the 
energy-level of dot $a$ and dot $b$ at the same level 
and makes the two dots symmetric
($\omega_a =\omega_b$ and $c_a=c_b$), 
\begin{equation}
\left(
\begin{array}{c}
a(t) \\ b(t)
\end{array}
\right)
=
\left(
\begin{array}{cc}
\cos (\omega_0 t) & -i\sin  (\omega_0 t) \\
-i\sin  (\omega_0 t) & \cos (\omega_0 t) 
\end{array}
\right)
\left(
\begin{array}{c}
a(0) \\ b(0)
\end{array}
\right) e^{-i\omega_a t}
\end{equation}
When the time-dependent phase is removed by the 
interaction picture, 
this solution shows that it includes 
the NOT operation in quantum computing, 
one of the basic single qubit operations.  
In particular, when the charge is localized 
in one of the coupled dots 
in the initial state($a(0)=1$, $b(0)=0$), we have:
\begin{equation}
a(t)=e^{-i\omega_a t} \cos \omega_0 t, \ \  
b(t)= -ie^{-i\omega_a t}\sin \omega_0 t. 
\end{equation}
This shows that the localized electron moves completely 
between dot $a$ and dot $b$ with a period, $\pi/(2 \omega_0)$. 
Thus we can show the charge transfer is realized 
when we apply the voltage at which the energy-levels 
of the initially isolated dot coincide, 
which also corresponds to the case 
where the energy-level of ground state and 
that of the excited state of the coupled dots 
approach most closely (hatched area in Fig.\ref{fig4}), 
in the time period of  $\pi/(2 \omega_0)$. 
Moreover, in the case of a different initial condition of 
charge distribution, $(|0 \rangle + | 1 \rangle )$ 
/$\sqrt{2}$ is realized as a static state. 

Time spent for the transfer of the charge in a coupled quantum dot 
is given as $\tau_{\rm \delta}^A=\pi/(2\omega_0)$ with: 
\begin{equation}
\omega_0 = \frac{4}{\hbar} \left( \frac{V_0-E}{V_0} \right) 
\frac{E}{1+Kl_w} e^{-K l_{d}} 
\end{equation}
where $l_w$ is an average width of the quantum dot (= 5nm),
$l_d$ (=1.5 nm) is the width of the tunneling barrier between the 
two dots, $K= \sqrt{2m(V_0 -E)/\hbar^2}$ and 
and $E$ is the energy of incident electron. 
We obtained $\tau_{\rm \delta}^A \sim$ 12 picosec in the case of 
 Fig.\ref{fig4}, which is longer than 
the time obtained above($\tau_{\rm \delta}^C$). 
This is because this time-dependent approach numerically  
identifies the exact results when 
the two dots are far apart\cite{Yariv}. 
In any event, this numerical mismatch never 
changes the physical aspects of the coupled dot system.  

Although the speed of the operation becomes faster as the tunneling 
barrier between the coupled dots in a qubit becomes thinner, the wave 
function of the qubit of thin tunneling barrier does not localize 
sufficiently. In the case of our calculation of SiO${}_2$, the criterion 
of the minimum thickness of the tunneling barrier is considered to be 
around 1nm, where the switching speed is estimated to be sub-picosecond. 
Below, the switching speed is also discussed in relation to the 
measurement time. 

\section{Discussion} \label{sec:discussion}
\subsection{Estimation of decoherence time}
Here we roughly estimate the decoherence time 
in quantum computation of the proposed coupled dots 
embedded in the SiO${}_2$ material based on the results 
by Leggett {\it et al.}\cite{Leggett}. 
During the quantum computation, 
the voltage between the source and drain is kept zero 
and there is no flow of the detecting channel current. 
The decoherence in this case is considered mainly 
to originate from the phonon environments. 
The SiO${}_2$ is a polar material and  the optical phonon 
mode ( $\sim$ 0.153 eV) will be the major dissipation mechanism 
of the high temperature and high energy region.  
Here, we consider the low temperature region where only 
the  acoustic phonons play the major role in the decoherence mechanism. 
The effects of this dissipative environment 
on the two-state system is treated by the infinite bath of 
harmonic oscillators of acoustic phonons 
(spin-boson Hamiltonian) where 
the interaction term between the two-state system and the acoustic 
phonons is derived from that of  
the amorphous SiO${}_2$ \cite{Garcia,Wurger}.  
The spectral function, $J(\omega)$, is given in the Debye 
approximation  as:
\begin{equation}
\frac{1}{2\pi \hbar}J(\omega)=
\frac{\gamma^2}{2\pi^2 \hbar \rho c^5 } \omega^3
+\frac{\gamma^2 \nu^2}{2\pi^2 \hbar \rho c^3 d^2} \omega, 
\label{eqn:spectral}
\end{equation}
where $\gamma \sim$ 10 eV, $c \sim$ 4300 m/sec, 
$\rho \sim$ 2200 kg/m${}^3$, $d \sim $ 0.5 nm, 
$\nu \sim 10^{-4}$ are a 
deformation potential, a sound velocity, a density, 
a lattice constant, and a dimensionless 
parameter, respectively. 
Here, we use the value of the deformation potential 
of the electrons in the bulk Si, 
because in the model of Ref.\cite{Garcia,Wurger}, 
the particle in the two-state system is assumed to be an atom. 
The first term of Eq.(\ref{eqn:spectral})
is the superohmic part and the second is the ohmic part. 
From Ref.\cite{Wurger}, the temperature where the ohmic part 
appears is estimated to be less than mK. 

First we estimate the decoherence time of the superohmic term. 
Here we treat the case of no bias and 
denote the bare tunneling frequency as $\Delta \equiv c_a=c_b$. 
According to Leggett {\it et al.}\cite{Leggett}, 
the two-state system without bias voltage 
shows the underdamped $coherent$ oscillation 
where the damping rate, $\Gamma_{\rm so}$, at $T=0$, is given as
\begin{equation}
\Gamma_{\rm so} =J_{\rm so} (\tilde{\Delta})/4\pi
= \frac{\gamma^2 \tilde{\Delta}^3}
{4 \pi \hbar \rho c^5},
\label{eqn:decoherence_rate}
\end{equation}
where $\tilde{\Delta}$ is the renormalized form of the 
bare tunneling frequency, $\Delta$, 
defined as 
\begin{eqnarray}
\tilde{\Delta}&=& \Delta \exp \left(-\frac{1}{2\pi \hbar}
\int_0^{\infty} d \omega \frac{J(\omega)}{\omega^2}
\right)
\nonumber \\
&=& \Delta \exp \left(-\frac{\gamma^2 \omega_c^2}{
2 \pi^2 \hbar \rho c^5} \right).
\label{eqn:delta}
\end{eqnarray}
When we take the above parameters of a-SiO${}_2$ of 
Ref.\cite{Garcia} and cutoff, 
$\omega_c = k_{\rm B} \Theta_D / \hbar$ with $\Theta_D \sim$ 450K, 
the value of the factor in the exponential is less than 
-10${}^{3}$, which extremely reduces the value of $\tilde{\Delta}$
( $\hbar \Delta \sim 10^{-5}$ eV in the above case). 
The decoherence time derived from the superohmic dissipation, 
$\tau_{\rm so} = 1/\Gamma_{\rm so}$, increases,  
as the $\tilde{\Delta}$ decreases.  
The direct calculation of the decoherence time becomes 
more than seconds. 
This will be because the true microscopic values 
will be different from those used above, 
which will be partly the same situation as Ref.\cite{Shore}. 
When we use bare tunneling frequency, $\Delta$ instead of 
$\tilde{\Delta}$ in Eq.(\ref{eqn:decoherence_rate}) 
in order to estimate the shortest decoherence time 
at the present, 
$\tau_{\rm so} \sim $ 4.8$\times 10^{-7}$ sec, during which 
thousands of quantum calculations can be realized 
in the proposed system where the one-step calculation 
is executed in a few picoseconds.  
Because $\Delta \ll \omega_c$, this underdamped behavior 
persists up to the finite temperature 
(see Leggett et al.\cite{Leggett}).
However, in order to show the numerical behavior at the finite 
temperature, we need the microscopic material values which 
appear in Eq.(\ref{eqn:delta}). Thus, we cannot show the maximum 
temperature of operation limited by the above superohmic term here. 

At low temperature region less than mK, ohmic dissipation 
should be considered. 
The dimensionless ohmic dissipation coefficient, $\alpha 
=\gamma^2 \nu^2/(2\pi^2 \hbar \rho c^3 d^2) \sim 
2 \times 10^{-7}$,  shows that in our case 
the {\it coherent} oscillations survive and 
the contribution of the incoherent part vanishes 
in this small $\alpha$ regime ( Ref.\cite{Leggett}).  

These 'long' decoherence times will originate from the 
high potential barrier(SiO${}_2$) between the two coupled 
quantum dots. 
By contrast, the short transition time via 
acoustic phonons in the two-state model of glasses is 
estimated to be of the order of $10^{-12}$ sec, 
which is derived from a low barrier height ($\sim 0.2$eV ) 
and a short distance of the two states ($\sim $ 0.1nm)
\cite{Anderson}. 
These 'long' decoherence times will be related with 
the 'phonon bottleneck' of Ref.\cite{Zanardi} and 
effects of phonons are different from those in the bulk\cite{Tsu1}.

The promising results of the above estimation 
of the 'long' decoherence time might be due to 
the simple two-state model of the ideal quantum  
dots apart from the question of the true microscopic values. 
When we consider the transitions to the excited 
energy-levels in each quantum dot
(about 0.018eV($\sim$ 210K) above the ground state 
in the 10nm Si quantum dot), 
which will occur at higher temperature regions
($>$ 210K), 
it is possible that the desirable quantum operation 
will be limited. 
Also, as the temperature rises, the effects of 
optical phonons cannot be neglected and the 
decoherence time will be reduced by the energy 
exchange between the electron and the optical 
phonon modes. 
In particular, when the quantum gate is operated in 
the AC gate voltage mode of high frequency in a general 
quantum operation, 
the dipole of the charge distribution in the coupled 
quantum dots will be more strongly affected by the 
electronic environment and the decoherence time 
will be reduced.  
Although there are many problems that need to 
be investigated in more detail, 
the above  results show that the quantum computing 
in the semiconductor coupled quantum dots seems to 
be realizable from the viewpoint of one of the elementary 
steps of the investigation\cite{Next}.

\subsection{Measurement process}
Next, the measurement process of the MOSFET structure 
is discussed in more detail. 
We treat the linear region of channel current, $I_{\rm d}$, 
as $I_{\rm d}= g_{\rm m} (V_{\rm G}- V_{\rm th})$, 
where $g_{\rm m}$ is a transconductance of the FET, 
and $V_{\rm th}$ is a threshold voltage. 
The change of the charge distribution in the coupled dots
induces a shift of the threshold gate voltage, 
$\Delta V_{\rm th}$, which is measured through the 
variation of the detecting channel current, $\Delta I_{\rm d}$, 
by the field effect.
The $\Delta V_{\rm th}$ of a single qubit is 
given approximately by 
$e d_{\rm ab}/\epsilon $ 
where $d_{\rm ab}$ is the distance between the two 
dots in a qubit, and found to be of the order of 
a few tens of meV.  This magnitude of the shift of 
the threshold voltage is found to be of a somewhat  
smaller order than that reported by Guo {\it et al.}\cite{Guo}.
The corresponding $\Delta I_{\rm d}$ is given by 
$\Delta I_{\rm d} =-g_{\rm m} \Delta V_{\rm th}$. 
This measurement process is considered to be similar to that 
of a quantum point contact(QPC). 
The measurement process of the coupled dots by QPC 
has been intensively investigated\cite{Gurvitz,Hackenbroich}. 
Here we use the results of Gurvitz\cite{Gurvitz}. 
The measurement time, $\tau_{\rm ms}$, at the channel current 
of $I_{\rm d}$, can be described as
\begin{equation}
\frac{1}{\tau_{\rm ms}}=
\left(\sqrt{\frac{I_{\rm d}+\Delta I_{\rm d}}{e}}
-\sqrt{\frac{I_{\rm d}}{e}}\right)^2. 
\end{equation}
The behavior of the detector 
can be classified depending on 
whether $1/\tau_{\rm ms} \ll \Delta$ or 
$1/\tau_{\rm ms} \gg \Delta$, where $\Delta$ is the  
tunneling frequency of the electron in a qubit. 
The first case, $1/\tau_{\rm ms} \ll \Delta$, is 
called  a 'weak damping' 
which implies that the electron oscillation 
in the qubit is faster than the detection 
and we will not be able to decide the position of 
the electron in a qubit. 
The second case, $1/\tau_{\rm ms} \gg \Delta$, 
is called  a 'strong damping' 
which implies that many electrons can flow through 
the channel region during an electron oscillation in a qubit. 
In the latter case, we can induce a 'Zeno time', 
$\tau_{\rm Z} \sim (1/\tau_{\rm ms})/(8\Delta^2)$, 
and observe the position of the electron in the 
time interval between $\tau_{\rm ms}$ and $\tau_{\rm Z}$ by 
the continuous measurements by Zeno effect. 

When we apply these arguments to the experiments 
of Guo {\it et al.}\cite{Guo} at $\Delta V_{\rm th} = 30$ meV, 
we have $g_{\rm m} \sim 1.8 \times 10^{-9} \Omega^{-1}$ 
and $\tau_{\rm ms} \sim 1.7 \times 10^{-6}$ sec $(\gg 1/\Delta$), 
if $\hbar \Delta =10^{-5}$ eV. 
This shows that the measurement parameters in the 
results of Ref.\cite{Guo} are not those of a good detection 
for $\hbar \Delta =10^{-5}$ eV. 
The $g_{\rm m}$ can be simply increased by increasing the bias between 
the source and drain\cite{Tiwari}.  These quantum-dot memories are 
considered to be prototypes, and, in the future, great improvement can be 
expected, for example, by reducing the resistive parasitics between the 
source and drain. However, when we would be unable to find a solution 
capable of improving the speed of transconductance by three orders, 
it would be necessary to reduce the speed of the gate operation, 
which can be controlled by the thickness of the tunneling barrier 
between the coupled dots, so as to realize better detection. In any 
events, the optimal speed of the gate operation will be able to be 
increased  as the developing fabrication technology improves the 
measurement speed. 

Lastly, we consider whether the SET  structure proposed 
by Shnirman {\it et al.}\cite{Shnirman} is 
more suitable for the measurement than the MOSFET structure. 
If we adopt the SET instead of MOSFET, 
the Josephson coupling energy, $E_{\rm J}$, 
of Shnirman {\it et al.}\cite{Shnirman} corresponds to 
our tunneling matrix element, $\hbar \Delta$,  and 
their charging energy term corresponds to our 
bias term, $E_a -E_b$, respectively. 
This is possible because the Hamiltonian of 
Shnirman {\it et al.}\cite{Shnirman} 
is described by the two-state system, 
and their model is considered to be universal in the 
measurement process in quantum computing. 
In this case we have to reduce the $E_{\rm set}$ to 
less than that of Shnirman {\it et al.} 
in order to suppress the bias term in the coupled dots 
and prevent the breakdown of coherence of the two-state system.  
Thus, 
the MOSFET structure seems to be available 
for the read-out device of the semiconductor qubit system,  
although the strict comparison will be needed. 
Moreover, problems may be 
ameliorated as the process 
technology of conventional LSI advances.

\subsection{Fabrication of the coupled quantum dots}
The coupled dots (qubits) 
can be fabricated by applying the self-limiting oxidation 
process of Si nanostructure\cite{Liu}. 
For example, 
the oxidation process 
after forming Si nanocrystals on a thin amorphous 
Si layer via SiO${}_2$ thin film changes 
the amorphous Si layer and leaves Si dots 
only under the top Si nanocrystals which also remain.  
Thus forming the coupled dot system is more feasible 
than controlling donor atoms in substrates. 
The small fluctuation of the dot sizes is not serious 
because the on/off gate voltage can be adjusted to be 
initialized depending on each energy-level of 
each qubit. 
Another concern is that interface traps may be 
another localized state and break coherence of the 
quantum calculation. 
However, the density of the trap state 
($\sim$ 10${}^{10}$ cm${}^{-2}$) is smaller 
than the assembly of the nanocrystals 
($\sim$ 10${}^{12}$ cm${}^{-2}$). 

\section{Conclusions} \label{sec:conclusion}
We have proposed a quantum computer 
based on the coupled dot system, 
which can be realized by developing the technology 
of the single-electron memory with Si nanocrystals. 
The basic idea is (1) the electron is localized 
in one of the asymmetric coupled dots. 
(2)The electron transfer takes place from one 
dot to the other when the energy-levels of the 
coupled dots are set close. (3)The Coulomb interaction 
between the coupled dots mutually affects the energy 
levels of the other coupled dots. 
The estimated decoherence time is found to 
permit a sufficient number of quantum calculations to be executed.  
The proposed system, where 
the direct combination of the quantum circuit and 
the conventional circuit is possible, 
is shown to be a promising candidate for the quantum computer. 


\acknowledgments
The author is grateful to K. Sato, N. Gemma, S. Fujita, K. Ichimura, 
K. Yamamoto, J. Koga and R. Ohba for fruitful discussion.

\begin{figure}
\caption{
Coupled quantum dot as qubit:
large quantum dot and small quantum dot are 
coupled such that the larger dot is set 
close to the channel from where one excess 
electron is inserted into the coupled dots. 
The smaller dot is set near the gate electrode via 
thick tunnel barrier which controls the 
energy-levels of the coupled dots.
The localized electron in a larger dot 
expresses the $|1>$ state and that 
in a smaller dot expresses the $|0>$ state.}
\label{fig1}
\end{figure}

\begin{figure}
\caption{
Quantum gates (controlled-NOT-gate) are constituted by setting 
the coupled dots of Fig.1 close to each other 
with the common channel. 
Solid lines show the path of electron tunneling. 
Dotted lines show the electric fields generated 
between quantum dots or between quantum dots and gates.}
\label{fig2}
\end{figure}

\begin{figure}
\caption{
An example of the $N$ coupled dot system of quantum computing. 
Dots are coupled in the longitudinal direction. 
The electron transfer in the lateral direction is assumed to be 
neglected. 
The FET channel structure enables the detection of the small 
signal of the charge distribution in coupled quantum dots.}
\label{example}
\end{figure}

\begin{figure}
\caption{Relation between the energy-levels of electrons 
of a target qubit 
and the gate bias for the cases in which the control 
qubit is in '$|1 \rangle$' state and '$|0 \rangle$' state. 
The structure is: channel/SiO${}_2$(2.5nm)/Si nanocrystals 
(6nm:dot $a$)/SiO${}_2$(1.5nm)/
Si nanocrystals (4nm:dot $b$)/SiO${}_2$(7nm)/Gate.  
$A^{(1)}$ and $B^{(1)}$  show the localized regions in gate bias 
for a larger dot (dot $a_1$ in Fig.2) and 
a smaller dot (dot $b_1$ in Fig.2), respectively, when 
the control qubit is in $| 1 \rangle $ state. 
$A^{(0)}$ and $B^{(0)}$  show similar regions 
when the control qubit is in $| 0 \rangle $ state. 
Hatched areas show the regions of the  delocalization 
where the wave functions spread over the two dots. 
This area shifts depending on whether the control qubit is in 
'$|1 \rangle$' state or '$|0 \rangle$' state. 
At the boundaries of these areas, 
wave functions are delocalized less than 98 \% in one of the dots 
and at their centers, $V_{\rm res}^{(1)}$ or 
$V_{\rm res}^{(0)}$, wave functions are 
equally distributed in both dots.}
\label{fig4}
\end{figure}

\begin{figure}
\caption{Spatial dependence of $|\psi|^2$ of the target qubit when 
the gate bias is $V_{\rm res}^{(1)}$: 
(a) the control qubit is in '$|1 \rangle$' state
(the charge of the control qubit is localized in a larger dot near 
the channel and the potential of the dot near the channel in 
target qubit is raised),
(b) the control qubit is in '$|0 \rangle$' state
(the charge of the control qubit is localized in a smaller dot 
and the potential of the smaller dot in 
target qubit is raised).
Wave functions are normalized in the lateral regions of the figures. 
The amplitude of the normalized wave function refers 
to the left scale and the potential profile of the 
qubit refers to the right one.
This shows the controlled-NOT operation in which 
the state of the target qubit is changed  
in a few picoseconds (dynamical properties) 
only if the control qubit is in $|1\rangle $ state. }
\label{fig5}
\end{figure}

\end{document}